\documentstyle [preprint, aps, multicol, epsfig]{revtex}
\tightenlines

\begin{document}

\bibliographystyle{prsty}

\draft

\title{Experimental Demonstration of Fermi Surface Effects at Filling
Factor 5/2}

\author{R.L. Willett,  K.W. West, L.N. Pfeiffer}

\address{Bell Laboratories, Lucent Technologies, Murray Hill, NJ 07974}

\maketitle

\begin{abstract}
Using small wavelength surface acoustic waves (SAW) on ultra-high mobility heterostructures,
Fermi surface properties are detected at 5/2 filling factor at temperatures higher than 
those at which the quantum Hall state forms.  An enhanced conductivity is observed at 
5/2 by employing sub 0.5 $\mu $m SAW, indicating a quasiparticle mean free path substantially 
smaller than that in the lowest Landau level.  These findings are consistent with the presence
of a filled Fermi sea of composite fermions, which may pair at lower temperatures to form the
5/2 ground state.\vspace{0.5in}
\end{abstract}

Since its discovery\cite{Willett:87}, the quantum Hall state at 5/2 filling factor 
(second Landau level, N=1) has remained enigmatic. The state violates the fundamental 
property of all other fractional quantum Hall states in that they occur at 
odd-denominator filling factors in order to preserve the antisymmetry of the  
many-particle wavefunction.\cite{Halperin:83}  Subsequent experiments\cite{Eisenstein:88}
using tilted B-fields to increase the Zeeman energy showed degradation of the 
quantum Hall effect at 5/2, which was taken as a possible sign of non-spin polarization in 
the ground state. While more recent experiments\cite{Pan:99} have confirmed 
the quantization of the Hall trace at 5/2, discovery\cite{Du:99,Lilly:99}
of exotic phase separated states in nearby higher Landau levels has shown that 
electron correlations manifest in numerous ways within the magnetic field range near 5/2. 
Just as tilted magnetic field can effect the anisotropic transport of the phase 
separated systems in high Landau levels ($N \geq 2$), it has been seen that tilted B-fields 
can likewise induce anisotropic transport effects at 5/2.\cite{Lilly2:99,Pan2:99}
As such the 5/2 state has experimentally demonstrated several peculiar properties 
consistent with its position in the magnetic field spectrum between the high 
Landau level stripe phases and the lowest Landau level (N=0) fractional 
quantum Hall states, which are understood in the picture of composite 
fermions.\cite{Jain:89,HLR:93} In this model of lowest Landau level physics 
the series of fractional quantum Hall states represent Landau levels for 
the quasiparticles, composite fermions, which near 1/2 are electrons and two associated 
flux quanta.  The applied magnetic field can leave the quasiparticle at zero 
effective magnetic field (just at 1/2), with 1/2 a compressible Fermi-liquid-like
state, or potentially at filled quasiparticle Landau levels (such as 1/3), which 
are incompressible quantum Hall states of the composite fermions.  As higher 
Landau levels are traversed, the validity of the composite fermion picture is 
questionable.

The theoretical description of the 5/2 state has been developed around 
models of paired composite particles.  Haldane and  Rezayi\cite{Haldane:88} 
derived a paired-electron state which is spin unpolarized, and so 
consistent with the early tilted-field experimental work, but also fit 
a pseudopotential profile that reflected the occurance of the state in 
the second Landau level.  This state can be considered a d-wave pairing 
of composite fermions, forming a spin-singlet.\cite{Read:00}  Moore and 
Read\cite{Moore:91} produced a spin polarized state developed as a 
Pfaffian that represents a p-wave pairing of composite fermions. The picture 
of a spin-polarized state at 5/2 at first appeared to be in conflict with 
the experiments using tilt, but gained support following numerical studies 
by Morf\cite{Morf:98} in which finite size systems were shown to have large overlap with 
the Moore-Read state. This work also showed that by strengthening the interaction, 
a transition to a Fermi-sea state could occur:  in-plane magnetic field can 
effect the interactions due to finite extent of the electron wavefunction out of the
2D plane.  Rezayi and Haldane\cite{Rezayi:00} used a different numerical approach 
and elaborated the results of Morf to describe the presence of a striped 
phase state, Fermi-liquid, or paired state dependent upon the interaction 
strength.  In this work they describe a condensation from Fermi-liquid 
to paired state at 5/2 for lower temperatures, but transition to striped 
phase for some in-plane field due to interaction changes.  This model 
shows a progression from Fermi-liquid at 1/2 (N=0) to paired state 
at 5/2 (N=1), to striped phase at 9/2, 11/2, 13/2, ...$(N \geq 2)$. 

While these theoretical pictures implicitly include the composite fermion 
Fermi-liquid as a precursor to the 5/2 state, no explicit experimental 
evidence to date has shown the existence of composite fermions at 5/2.
In the lowest Landau level at 1/2 numerous measurements have established
the existence of a filled composite fermion Fermi sea.  Surface acoustic 
wave studies\cite{Willett:90} first demonstrated an anomalous enhanced 
conductivity at 1/2 which was later recognized as ballistic transport of the 
quasiparticles over the small dimension of the SAW potential.\cite{Willett:93a}
Further SAW measurements showed geometric resonance of the composite 
particle trajectories with the SAW\cite{Willett:93b} which allowed 
precise extraction of the sea's Fermi wavevector.  Other measurements 
supported this finding, again all applying a small length-scale
conductivity measurement: antidots\cite{Kang:93}, magnetic focusing\cite{Smet:96} 
and later resonance with a 1d line array.\cite{Smet:99} These measurements concerned
the state at 1/2, which demonstrates a robust Fermi sea over a wide 
range of temperatures, large range of electron densities, and in relatively 
low mobility electron systems, mobilities much lower than those necessary to 
support a strong 5/2 state or the higher Landau level stripe phases.  While 
Fermi surface properties were likewise readily extracted at 3/2 filling 
factor\cite{Willett:97} using these experimental methods, no evidence for 
Fermi sea formation at 5/2 was observed.

In this letter we present surface acoustic wave measurements that distinctly
demonstrate features at 5/2 filling factor which, in similarity to findings 
at 1/2, indicate the presence of a composite fermion Fermi surface.  These 
SAW propagation features at 5/2 are observed at temperatures above those
at which the system forms a quantized Hall state, and were derived using 
samples of particularly high mobility.  These results required SAW of much
smaller wavelength than those used to observe Fermi surface effects at 1/2, 
implying the composite particle mean-free-path at 5/2 is substantially 
shorter than that at 1/2.  The enhanced conductivity at 5/2 for small 
$\lambda $ SAW has a magnetic field extent that may 
indicate a fully spin polarized Fermi sea in the second Landau level.
These results may be supportive of the model of composite fermion pairing at 5/2
in which quasiparticles condense into a Moore-Read like state.

In order to determine the presence of a Fermi surface at 5/2 we have examined
the small length scale conductivity using surface acoustic waves.  In a proposed 
composite fermion system the conduction will occur in zero effective magnetic
field at the filling factor value where the Chern-Simons field and applied field
are equivalent, in this case potentially at 5/2.  Away from 5/2 the charge will follow a 
cyclotron radius according to $R_{c} = \hbar k_{F}/e\Delta B$ where $k_{F}$ is
the composite particle's Fermi wavevector and the effective magnetic field is
$\Delta B = B_{applied} - B(\nu =5/2)$.  As effective magnetic field is increased,
the charged particle will ultimately follow a small orbit cyclotron path with guiding
center orthogonal to the driving E-field.  Conduction paths at zero and low 
effective magnetic field will preserve their trajectories for length scales smaller
than the scattering length; the charge transport will be ballistic.  It is these
ballistic transport phenomena\cite{Willett:97} near zero effective magnetic 
field that have been exploited in SAW, split gate, antidot, and focusing 
measurements to expose the composite fermion Fermi surface at $\nu $ =1/2.

These probes all apply a conduction measurement over some small length-scale which
must be smaller than the mean-free-path of the composite particle in order to expose
the consequences of zero effective magnetic field.  In the case of the surface 
acoustic waves, a longitudinal surface sound wave is launched across the 2DES and
its change in sound velocity is measured.  Since GaAs is piezoelectric the wave has 
an associated E-field applied over the wavelength of the SAW and in the propagation
direction; the electron system responds to this E-field. The sound wave is slowed and 
attenuated through this interaction, with the sound velocity altered according to 
 $\Delta v/v$ = $\alpha /( 1 + ( \sigma_{xx}/ \sigma_{m})^2)$, $\alpha $ the piezo-electric
coupling, $\sigma_{m} $ determined by the sample parameters, and $\sigma_{xx} $ the sheet
conductivity at that SAW frequency and wavevector.\cite{Wixworth:86}  This relaxation 
response shows heuristically that a minimum in $\sigma_{xx} $ causes a peak in $\Delta v/v$,
so that quantum Hall states display peaks in the sound velocity shift.  However, in such
a SAW measurement at zero effective magnetic field for a composite fermion the 
resulting $\Delta v/v$ can be anomalously small, forming a local minimum, as was shown for the composite 
fermion system at 1/2 filling factor.  If the SAW wavelength is smaller than the
quasiparticle mean-free-path, the quasiparticle conducts ballistically across
the acoustic wave, displaying a minimum in $\Delta v/v$  (enhanced conductivity)
at that filling factor compared to the the ultrasound response for the surrounding B-field 
range, or for longer wavelength SAW.  It is this effect that is an indication of Fermi
surface formation, as documented at 1/2 filling factor.  The larger the quasiparticle 
mean-free-path, the larger the SAW wavelength that may be used in order to show 
enhanced conductivity.  This enhanced conductivity grows as the SAW wavelength 
is reduced below the quasiparticle mean-free-path,  with 
$\sigma (q)  \sim q_{SAW} $, $ q_{SAW} = 2 \pi/ \lambda _{SAW}$.\cite{HLR:93}

In this study we examine the 2DES at 5/2 filling for just such enhanced conductivity.
In this work we optimize this search by applying low insertion loss small wavelength SAW to the 
highest purity heterostructures available, with potentially large quasiparticle 
mean-free-paths.  A series of six samples from a single wafer were used in this
study.  Mobilities for samples from this wafer are in excess of 28$\times$10$^{6}$cm$^{2}$/V-sec, 
substantially larger than sample mobilities used in previous measurements
addressing 1/2 composite fermions.  Measurements were performed in a He3 refrigerator,
with particular accomodation of low-loss high frequency lines.

The essential finding of this paper is demonstrated in Figure 1, which 
shows SAW response at 5.8GHz and d.c. tranport for a high mobility 
sample at 280mK.  The d.c. transport demonstrates the shallow minimum 
at 5/2 that is characteristic of this state at temperatures higher than
those at which the quantum Hall effect is observed.  Such a minimum 
in resistivity ($\sigma_{xx} $) should be reflected in the ultrasound 
response as a maximum at 5/2.  Instead a distinct minimum 
has formed at 5/2 in $\Delta v/v$, which is as expected for the 
enhanced conductivity of a composite fermion system.  This effect 
at 5/2 is substantially smaller than that shown at filling factor 3/2.

To further demonstrate this signature of the presence of composite fermions a 
range of SAW wavelengths (frequencies) were used. 
As noted previously the enhanced conductivity (minimum at 5/2) 
should increase as smaller wavelengths are used in measurement.  From Figure 2 
this trend is readily apparent.  No obvious minimum is observed for 2.1~GHz 
SAW (1.3 $\mu $m).  At 4~GHz the local maximum in $\Delta v/v$ expected from 
d.c. response is distorted. At 5.8~GHz (0.48 $\mu $m) a minumum at 
5/2 is clearly formed, and a substantial minimum is present using 7.8~GHz SAW. 
This progression to larger effect at 5/2 with increasing SAW wavevector 
is just as that observed at filling factor 1/2 but in that case using
significantly longer wavelengths.  From our observations 
one can state that the enhanced conductivity is first manifest for SAW 
frequencies between 4 and 6 GHz, or a wavelength of about 0.6 $\mu $m .  
The onset of enhanced conductivity for 1/2 filling factor (in lower 
mobility samples) was roughly 500~MHz, suggesting that the mean free path 
for the 5/2 composite fermion is almost an order of magnitude smaller 
than that of the 1/2 composite fermion.

The crude temperature properties of this effect at 5/2 are shown in Figure 3.  As the 
temperature is increased to near 1K the minimum in $\Delta v/v$ is lost, which is at a
much lower temperature than for the composite fermions at 1/2 filling factor 
($\sim 4K$)\cite{Willett:93a}.  The Fermi wavevector for the 5/2 composite 
fermion system is expected to be dependent upon a charge density that is 
five times less than the electron system due to
the inert filling of the lower Landau level.  Subsequently thermal broadening
is imposed upon a relatively smaller Fermi wavevector.  An important 
measurement is the evolution of the $\Delta v/v$
response at 5/2 as the temperature is lowered, as it is here that the transition from
filled Fermi sea to quantum Hall fluid should be observed.  This measurement was not
possible given our available temperature range, since the quantum Hall effect is observed
at 5/2 in these samples at temperatures less than about 80mK.

By examining the magnetic field width of the enhanced conductivity effect 
at 5/2 the spin population can be preliminarily addressed. The width of the $\Delta v/v$ 
minimum is determined by the composite particle cyclotron radius, which is 
proportional to the Fermi wavevector, which in turn is determined by the 
charge density $n^{*}$ filling the Fermi sea: ($ k_{F} = ( 4 \pi n^{*} )^{1/2}$).  
For a first look at this question of spin population 
the width at 5/2 can be compared to that of the 3/2 effect, which by previous 
studies\cite{Willett:97} was shown in geometric resonance effects to
be a fully spin polarized Fermi sea\cite{note:01}. Note that while 
SAW and antidot resonance measurements\cite{Kang:01} may be consistent 
with a fully spin polarized Fermi system, activation energy 
measurements\cite{Du:95} indicate that 3/2 is not spin polarized: 
this discrepancy remains unresolved.  With proper consideration 
of the lower Landau level filling, the width at 5/2 should 
be about 0.45 that of the 3/2 width.  
From the raw magneto-acoustic data of Figure 4 and other measurements 
summarized in the inset, the preliminary indication 
is that the width of the effect at 5/2 is consistent 
with this assessment, both for data at 7.8 and 5.8 GHz.   
These data suggest that 5/2 and 3/2 are similar in spin polarization 
as measured by SAW.  If it is the case that 3/2 is spin polarized,
this implies that the precursor composite fermion Fermi sea at 
5/2 is spin polarized. It is reasonable that this spin
population could be sustained upon cooling as the system pairs and condenses into
a ground state such as described by Moore and Read.\cite{Moore:91}
This qualified result is subject to further studies of the 5/2 width.

Interesting details of the ultrasound response are suggested 
by the high frequency (7.8~GHz) magnetic field sweep measurements in 
Figures 2 and 4.  The minimum in $\Delta v/v$ at 5/2 in each case 
appears to have some structure: they are not simple minima as were observed 
in early studies of 1/2 filling factor\cite{Willett:90,Willett:93a}. 
Other preliminary measurements not shown here corroborate the 
finding of structure within the effect 
at 5/2, but only for data taken at our lowest available temperatures.  
A geometric resonance of the cyclotron orbit with 
the surface acoustic wave could produce such structure, 
however this is not supported by 
the data here since at 3/2 such resonance effects are not present.  
Instead, the structure at 5/2 may indicate the partial formation 
of some quantum Hall liquid within the path traversed by the surface wave  
but not apparent in the d.c. transport. Further 
examinations of these possibilities are underway.

In conclusion, our findings demonstrate a minimum in sound velocity 
shift at 5/2 filling factor at temperatures higher than those at 
which the quantum Hall effect is observed, and only for small SAW 
wavelengths.  This effect, which becomes more pronounced at smaller 
SAW wavelength, is consistent empirically with the past findings at 
1/2 filling factor indicating the presence of a composite fermion 
Fermi surface.  The wavelengths needed to reveal this effect
are substantially smaller than those in the 1/2 studies, implying 
the 5/2 quasiparticles have a significantly smaller mean-free-path.
Our finding of Fermi surface effects at 5/2 is consistent with 
theoretical pictures describing the pairing of composite fermions which then 
condense into a ground state, which in turn demonstrates a quantum Hall effect at 
5/2. 

We would like to acknowledge useful discussions with S. Simon and N. Read.\vspace{0.5in}

%% put references here...  

\pagebreak

FIGURE CAPTIONS\vspace{1in}

Figure 1.  Surface acoustic wave sound velocity shift versus magnetic field in a high mobility  
$(> 28\times 10^{6} cm^{2}/V-sec)$ 2D heterostructure (sample A) at T=280mK.  The SAW 
frequency is 5.8 GHz, corresponding to a wavelength of about 0.5~$\mu$m .  At 5/2 long wavelength 
SAW and/or low mobility samples show a maximum, rather than the local minimum shown here, 
which indicates enhanced conductivity at the SAW wavelength and frequency.\vspace{0.5in} 

Figure 2.  Sound velocity shift for a series of frequencies (wavelengths) demonstrating 
appearance of the enhanced conductivity for decreasing wavelength. All measurements were 
performed at 280mK, on samples A, C, D, and E.\vspace{0.5in}

Figure 3. Coarse temperature dependence of the enhanced conductivity at 5/2 for probing 
SAW at 5.8 GHz, using sample B, demonstrating that the local minimum at 5/2 is lost at 
high temperatures.\vspace{0.5in}

Figure 4. Comparison of the enhanced conductivity magnetic field width at 5/2 and 3/2.  The 
magneto-acoustic trace is taken using 7.8 GHz SAW at 280mK on sample F. The inset plots 
the width as derived in the magnetic field trace for different SAW frequencies at both 
5/2 and 3/2.  The solid horizontal lines through the 5/2 data in the inset plot are the 
width predicted for 5/2 if the spin population is similar to that of 3/2.\vspace{0.5in}

\pagebreak
\begin{figure}
\epsfclipon
\epsfxsize=12cm
\epsfbox{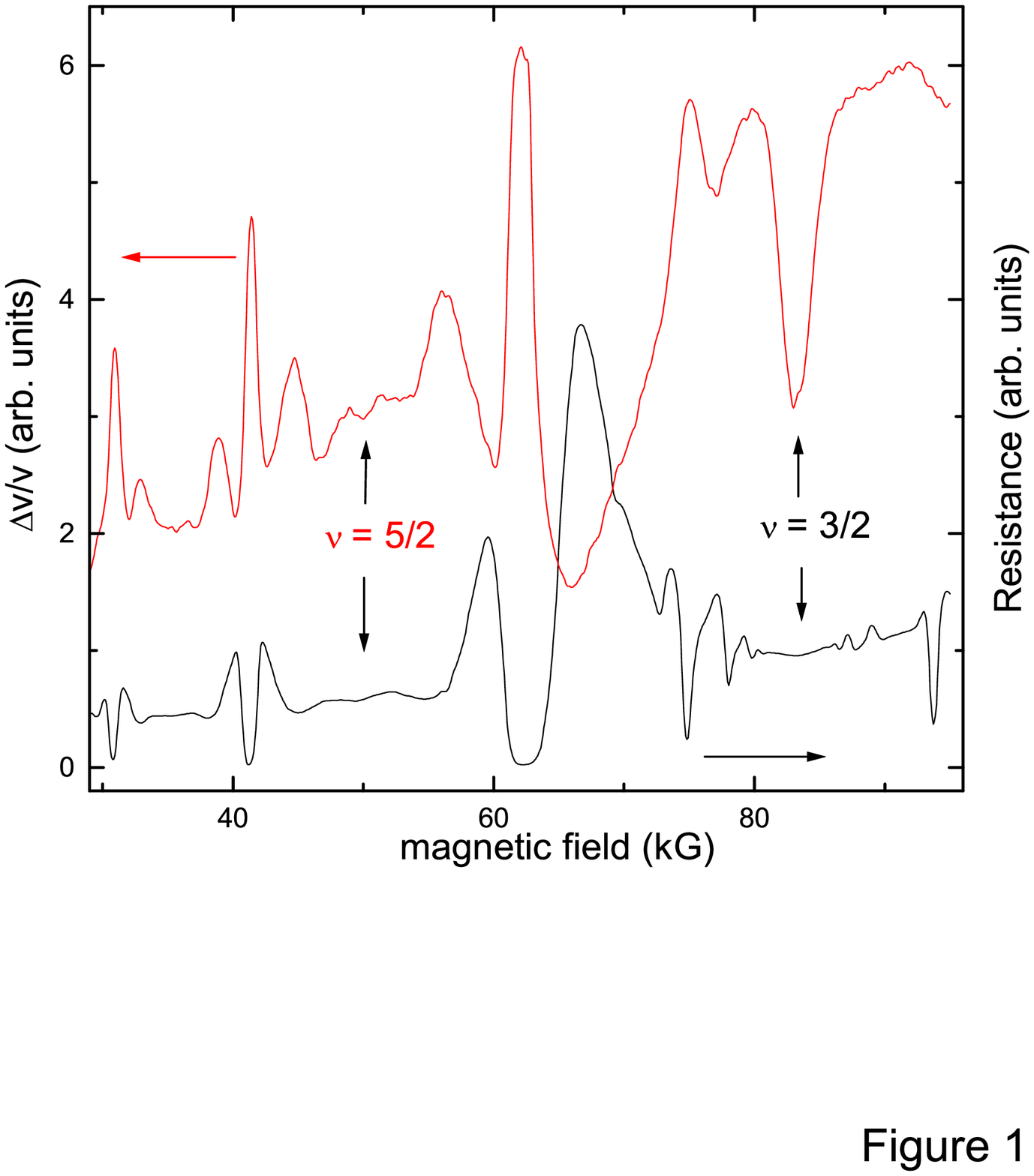}
\end{figure}

\pagebreak
\begin{figure}
\epsfclipon
\epsfxsize=12cm
\epsfbox{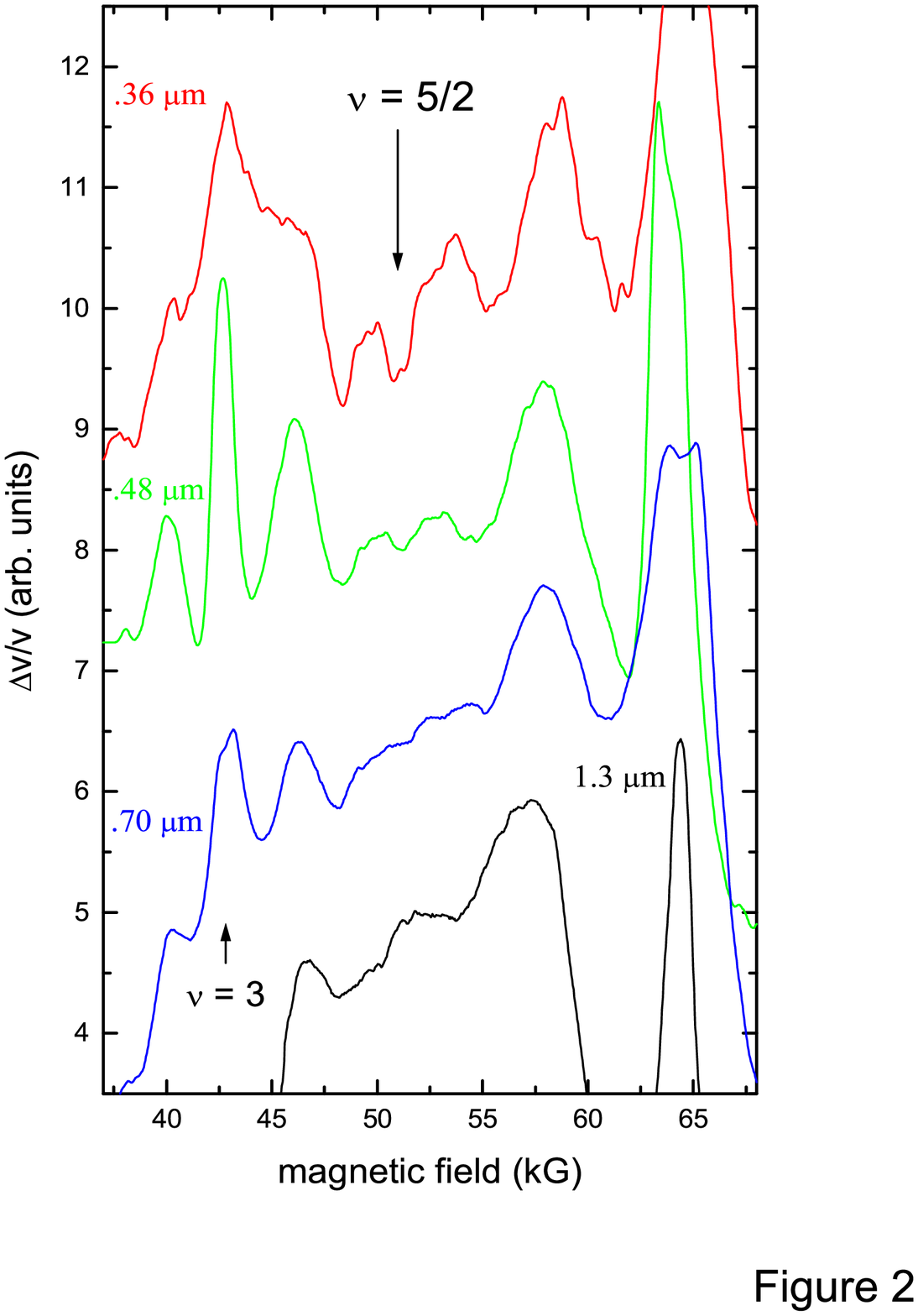}
\end{figure}

\pagebreak
\begin{figure}
\epsfclipon
\epsfxsize=12cm
\epsfbox{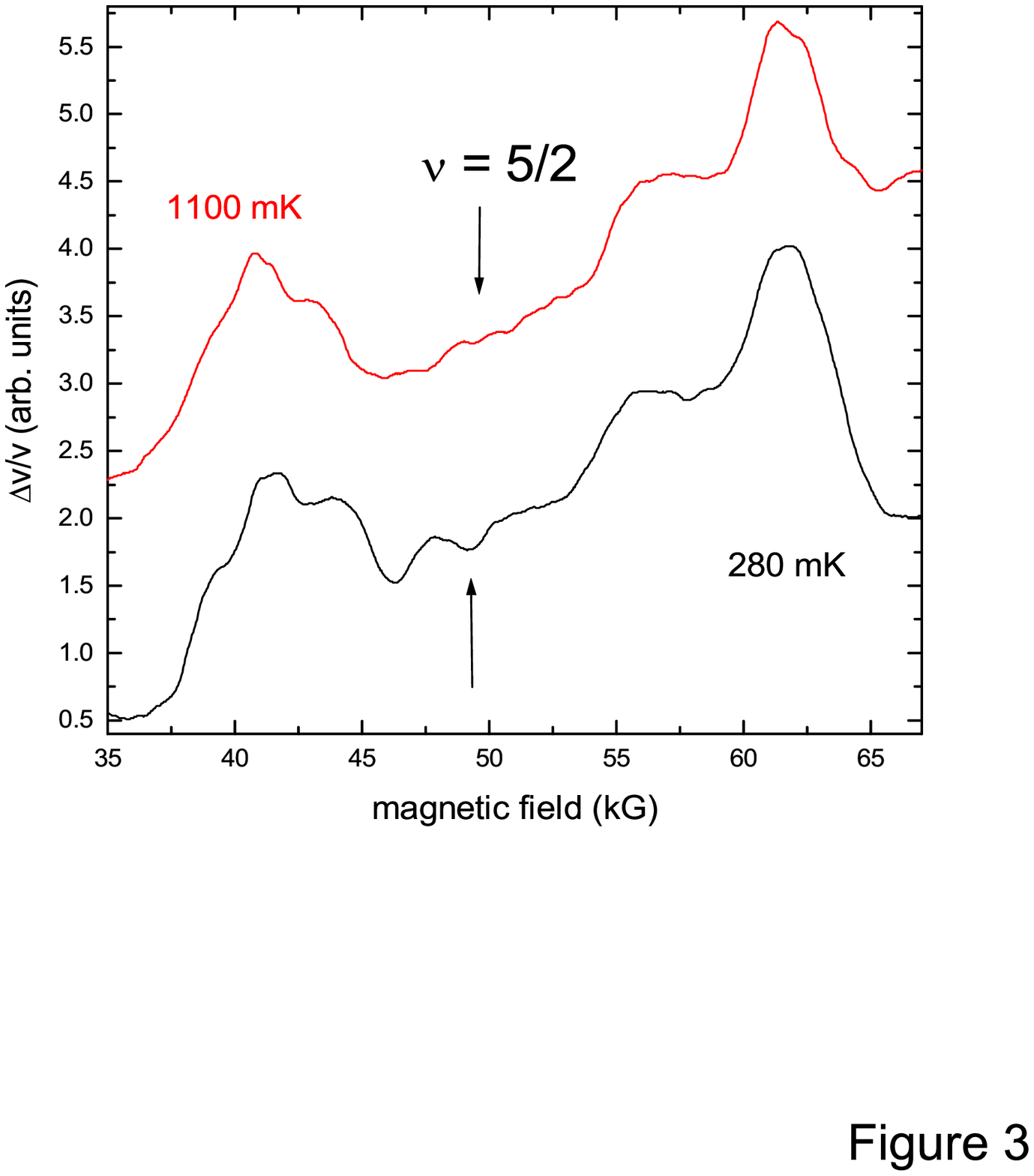}
\end{figure}

\pagebreak
\begin{figure}
\epsfclipon
\epsfxsize=12cm
\epsfbox{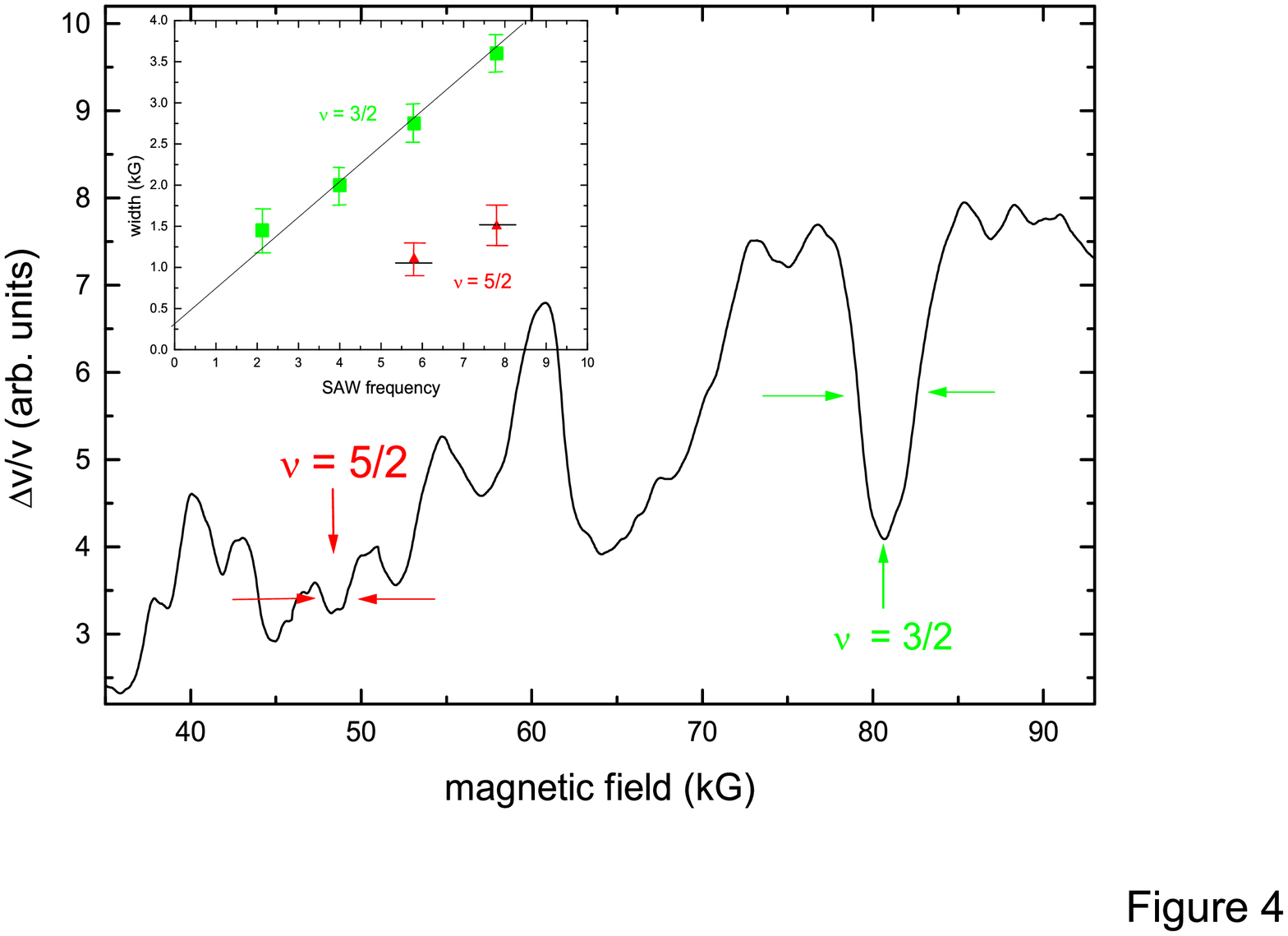}
\end{figure}

\end{document}